\documentclass[traditabstract]{aa} % for the abstract without structuration 
\usepackage{graphicx}
\usepackage{txfonts}
\usepackage{natbib}
\usepackage[bookmarks=false]{hyperref}

\newcommand {\hi} {{\rm H}\,{\small\rm I}}
\newcommand {\kms} {\,{\rm km\,s}^{-1}}

\newcommand {\kpc} {\,{\rm kpc}}
\newcommand {\Mpc} {\,{\rm Mpc}}
\newcommand {\de}{^{\circ}}
\newcommand {\mo}{\,{\rm M}_\odot}

\newcommand {\gsim}{\,\lower.7ex\hbox{$\;\stackrel{\textstyle>}{\sim}\;$}}
\newcommand {\lsim}{\,\lower.7ex\hbox{$\;\stackrel{\textstyle<}{\sim}\;$}}
\newcommand {\loB}{\,{\rm L}_{\odot, \rm B}}

\newcommand {\loS}{\,{\rm L}_{\odot, 3.6\,\mu{\rm m}}}
\newcommand {\loSpc}{\,{\rm L}_{\odot, 3.6\,\mu{\rm m}}\,{\rm pc^{-2}}}
\newcommand {\ml}{M/L}

\newcommand {\mlS}{M/L_{\rm 3.6\, \mu{\rm m}}}
\newcommand {\magasec}{\,{\rm mag}\,{\rm arcsec^{-2}}}

\begin{document}
  \title{A tale of two galaxies: light and mass in NGC\,891 and NGC\,7814}

  \subtitle{}

  \author{Filippo Fraternali
         \inst{1, 2}
         \, 
         Renzo Sancisi
         \inst{2, 3}
	  \and
	  Peter Kamphuis
	  \inst{2, 4}
      }

 \institute{
           Astronomy Department, University of Bologna, via Ranzani 1, 40127 Bologna, Italy\\
           \email{filippo.fraternali@unibo.it}
           \and
           Kapteyn Astronomical Institute, Postbus 800, 9700 AV, Groningen, The Netherlands
           \and
           INAF - Astronomical Observatory of Bologna, via Ranzani 1, 40127 Bologna, Italy
	    \and
	    Astronomisches Institut, Ruhr-Universit\"{a}t Bochum, Universit\"{a}tsstrasse 150, D-44801 Bochum, Germany 
         }

  \date{}

% \abstract{}{}{}{}{} 
% 5 {} token are mandatory

\abstract{
The two edge-on galaxies NGC\,891 and NGC\,7814 are representative of
two extreme morphologies: the former is disk-dominated while the
latter is almost entirely bulge-dominated. 
It has been argued (van der Kruit 1983) that since the two galaxies, 
which are optically so different, have similar rotation curves their 
total mass distributions cannot be related in any way to the light 
distributions. 
This would lead to the conclusion that dark matter is the dominating 
component of the mass.
We have derived new rotation curves from recent, high-sensitivity \hi\
observations and have found that the shapes of the rotation curves are
significantly different for the two galaxies. 
They indicate that in NGC\,7814 the mass is more concentrated to the
centre as compared to NGC\,891. 
This reflects the distribution of light which is more centrally
concentrated in NGC\,7814 than in NGC\,891.
Mass and light do seem to be closely related. 
This is confirmed by the analysis of the rotation curves in mass 
components: solutions close to the maximum light (bulge $+$ disk) do 
provide excellent fits. 
In NGC 891 bulge and disk can explain the rotation curve without any need 
for dark matter out to $\sim 15 \kpc$. 
In NGC 7814 the bulge dominates in the inner parts;
further out the rotation curve is well reproduced by a maximum disk but its 
$\ml$ ratio is excessively high. 
A substantial dark matter contribution, 
closely coupled to the luminous component, seems, therefore, necessary.
}
%   {1}
%   {2}
%   {3}
%   {4}
%   {5}


  \maketitle
%
%________________________________________________________________

\section{Introduction}

The distribution and relative importance of luminous and dark
matter in galaxies are still a matter of debate. The distribution of
mass in a spiral galaxy is inferred from its rotation curve. 
In the past three decades, rotation curves have been derived for a
number of spirals of various masses and morphological types. For the
interpretation of the rotation curves in terms of mass distribution
use has been made of multicolour photometry. The conclusion has been
that in the outer parts of galaxies there is a significant
discrepancy between the observed curve and the curve predicted from
the photometric and gas profiles.
Such discrepancy is usually interpreted as evidence for
the presence of dark halos around spiral galaxies. For the bright
inner parts of the disk, inside $R_{25}$, there is no such consensus. There
is a range of possibilities from ``maximum disks'' \citep{kalnajs83,
 vanAlbada86, kent86} with constant values of the $\ml$ ratio
(implying that luminous matter dominates) to ``minimum disks'' with
dark matter being the dominant component everywhere. 

\begin{figure*}
 \centering
 \includegraphics[width=0.9\textwidth]{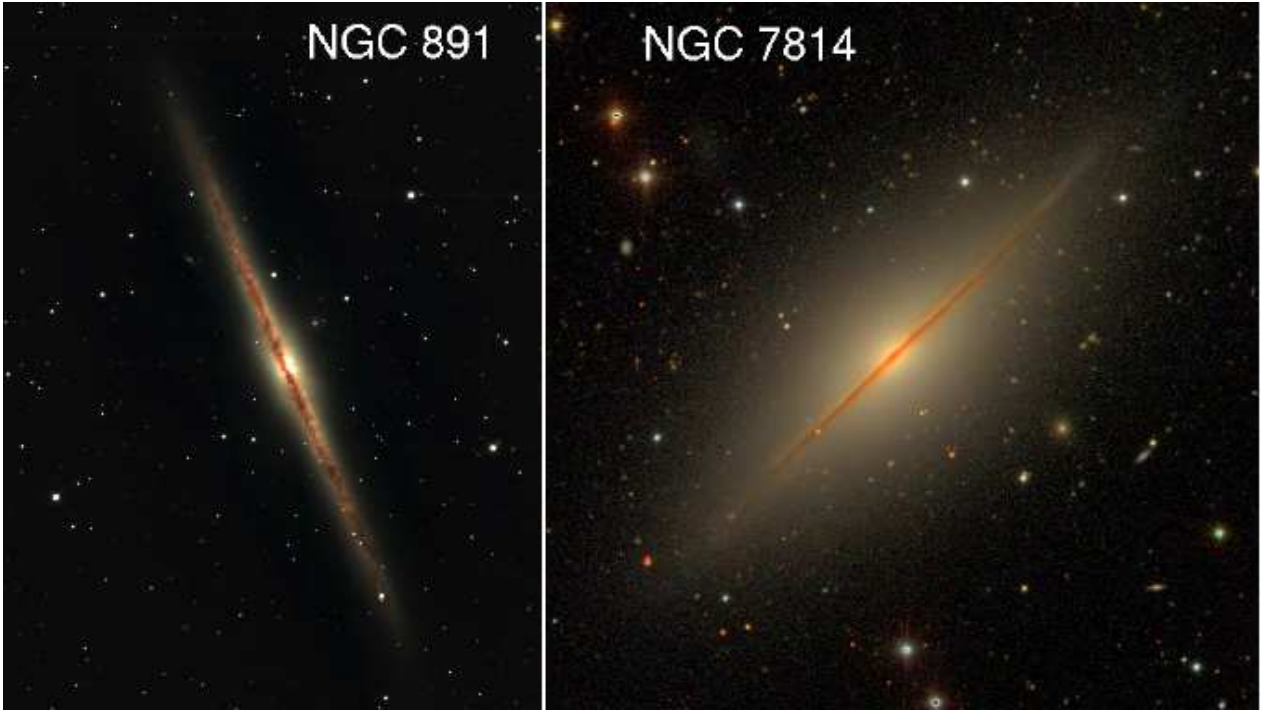}
 \caption{{\it Left:} Multi-band (Y, J, K) image of NGC\,891 obtained
   with the WIRCam at the Canada-France-Hawaii Telescope (CFHT). {\it
     Right:} Multi-band (g, r, i) image of NGC\,7814 from the Sloan
   Digital Sky Survey.}
\label{optical}
\end{figure*}

The debate on the relative distribution of 
luminous and dark matter in spiral galaxies is still open. 
The comparison between the two edge-on galaxies NGC\,891 and NGC\,7814 
provides a good illustration.
These two spirals are representative of two extreme morphologies: a
disk-dominated NGC\,891 and an almost entirely bulge-dominated
NGC\,7814 (see Fig.\ \ref{optical}). 
Van der Kruit (1983, 1987, 1995) pointed out that, in spite of this
striking difference in their light distributions,  these two galaxies
have, beyond the central $3-4 \kpc$, ``essentially identical'' rotation
curves.   
These curves were measured in the \hi\ 21-cm line but
they were not known inside 1 arcmin ($3-4 \kpc$) because \hi\ emission
was not detected there.
The conclusion was that ``if the distribution of luminous
matter were in any way related to the total mass distribution it would
not be possible for the disk-dominated and the bulge-dominated
galaxies to have such similar rotation curves''.  
If the two galaxies had indeed rotation curves of such similar shape
this would be contrary to the rule that in spiral galaxies there is a
close correlation between light distribution and rotation curve shape
\citep{sancisi04}. 
The contrast would also exist with non-Newtonian theories of gravity
which  would predict grossly different rotation curves
\citep{vdkruit95}. 
There is little doubt that the light distributions of NGC\,891 and NGC\,7814 
are very different.
But, do their rotation curves have indeed essentially identical shapes?

The rotation curves used by Van der Kruit for NGC\,891
\citep{sancisi79} and for NGC\,7814 \citep{vdkruit82} were obtained
from \hi\ observations with the Westerbork Synthesis Radio Telescope
(WSRT).  
The rotation curve of NGC\,891 was obtained directly from the
observations using a velocity ``envelope'' method, that of NGC\,7814
by constructing models.  
Both curves appear to be flat to a first approximation, but the \hi\
observations, especially for NGC\,7814, had poor signal/noise ratios
in the central regions and the curves had large uncertainties or could
not be derived at all. 

In recent years, new \hi\ observations were obtained for these two 
galaxies with better sensitivity with the WSRT.
From these observations we derived new rotation curves. 
We used Spitzer $3.6\, \mu{\rm m}$ observations to derive the photometric
parameters of the bulge and the disk components and perform
mass decompositions.
We were able to reach new conclusions on the relative
distribution of luminous and dark matter, which differ from those
mentioned above.

\begin{table}
\caption{Optical and \hi\ parameters for the two galaxies}
\label{tPars} 
\centering    
\begin{tabular}{lccc}
\hline\hline             
Parameter                         &     NGC\,891      &     NGC\,7814 & Ref.\\ 
\hline
Morphological type                &      Sb/SBb       &     Sab       & 1, 2, 3\\
Centre ($\alpha$ J2000) & $2^{\rm h}22^{\rm m}33.41^{\rm s}$& $0^{\rm h}03^{\rm m}14.89^{\rm s}$ &    \\
~~~~~~~~~~~~($\delta$ J2000)& $42^\circ20^\prime56.9^{\prime\prime}$&$16^\circ08^\prime43.5^{\prime\prime}$&   \\
Distance (Mpc)                    &          9.5      &     14.6      & 1$^a$\\
$L_{\rm B}$ ($\loB$)               & $2.5\times10^{10}$&$1.3\times10^{10}$&3$^b$\\
$L_{\rm K}$ ($\loB$)               & $7.8\times10^{10}$&$6.4\times10^{10}$&4\\
R$_{\rm 25}$ in B-band (kpc)    &          18.6      &  9.6      & 3, 5 \\
R$_{\rm 20}$ in K-band (kpc)     &          10.4      &     7.2      & 4 \\
Systemic velocity (km~s$^{-1}$)    & $528 \pm 2$       &   $1043\pm 4$ & 6, 7 \\
Total \hi\ mass ($\mo$)          & $4.1\times 10^9$  &$1.1\times 10^9$& 6, 7 \\
H{\sc\ I} inclination ($^{\circ}$) & $\gsim$89         &  $\sim90$     & 6, 7 \\
1 arcmin $=$ (kpc)                & 2.76              &  4.25         & 7 \\
\hline
\end{tabular}\\
\begin{flushleft}
(1) \citet{vdkruit81},
(2) \citet{garcia95},
(3) \citet{deVaucouleurs91},
(4) \citet{jarrett03},
(5) \citet{sdss},
(6) \citet{Oosterloo+07},
(7) this work.\\
$^a$ More recent determinations of the distances of NGC\,891 and NGC\,7814 using the tip of the red giant branch are fully consistent with these values, i.e.\ $9.1 \pm 0.4 \Mpc$ and $14.8^{+0.7}_{-0.6} \Mpc$ respectively \citep{Radburn-Smith+11}.\\
$^b$ Corrected for internal and Milky Way extinction. \\
\end{flushleft}
\end{table}

\section{Derivation of the rotation curves}

For NGC\,891 we used the \hi\ observations by \citet{Oosterloo+07} and for NGC\,7814 
those obtained by \citet{kamphuis08}. The latter consist of 4x12 hours integration with the WSRT (September 2004). The reduction was done with the MIRIAD package and standard calibration was applied. The result is a data cube consisting of 160 channels 
with a spatial resolution (FWHM) of $54.5" \times 12.9"$ and with a velocity spacing of 4.12 $\kms$. 

The derivation of the rotation curves was done in two steps. 
A first estimate was obtained by taking the envelope on the high rotational velocity side of the position-velocity diagram along the major axis. 
For a description of the method and a discussion of the uncertanties involved see \citet{sancisi79}.
The rotational velocity was derived by fitting Gaussian functions with fixed dispersions to the high rotational velocity sides of the line profiles (i.e.\ the lowest radial velocities on the approaching side and the highest on the receding one).
It was assumed that the gas is in circular motion, that there is gas emission at the line of nodes and that the velocity dispersion is constant and equal to $8\kms$.
Subsequently this rotation curve was used, together with the observed radial \hi\ density profile, as input for the modelling of the data cube.
This was done by assuming concentric rings. 
Each ring has its own rotational velocity and gas density. 
The inclination angle is fixed to $90\de$, determined by modelling the \hi\ datacube.
The position angle was measured using the total \hi\ map and the Spitzer data, the values agreed within $<0.5 \de$.
There is no evidence for a change of more than $2\de$ in position angle and $4 \de$ inclination out to the outer radius from which the rotation curve is derived \citep[for NGC\,891 see also][]{Oosterloo+07}.
The effect on the rotation curves of these changes would be less than $1 \kms$.

\begin{figure}
 \centering
 \includegraphics[width=0.5\textwidth]{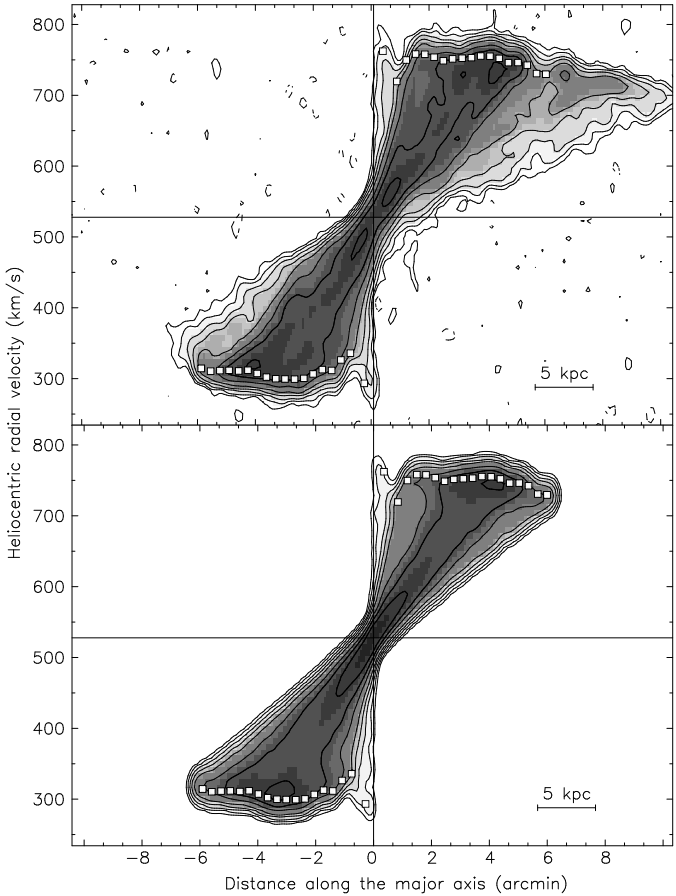}
 \caption{$Upper$: Position-velocity diagram along the major axis of the edge-on galaxy NGC\,891. The squares show the rotation curves derived separately for the approaching and the receding side. The contour levels are: 0.2, 0.45, 1.0, 2.0, 4.0, 10, 20, 40 ${\rm mJy}\,{\rm beam}^{-1}$.
$Lower$: Position-velocity diagram along the major axis for the model galaxy.
The model has been constructed for the symmetric part of the disk. 
1 arcmin$=2.76 \kpc$.}
 \label{pv891}
\end{figure}

\begin{figure}
 \centering
 \includegraphics[width=0.5\textwidth]{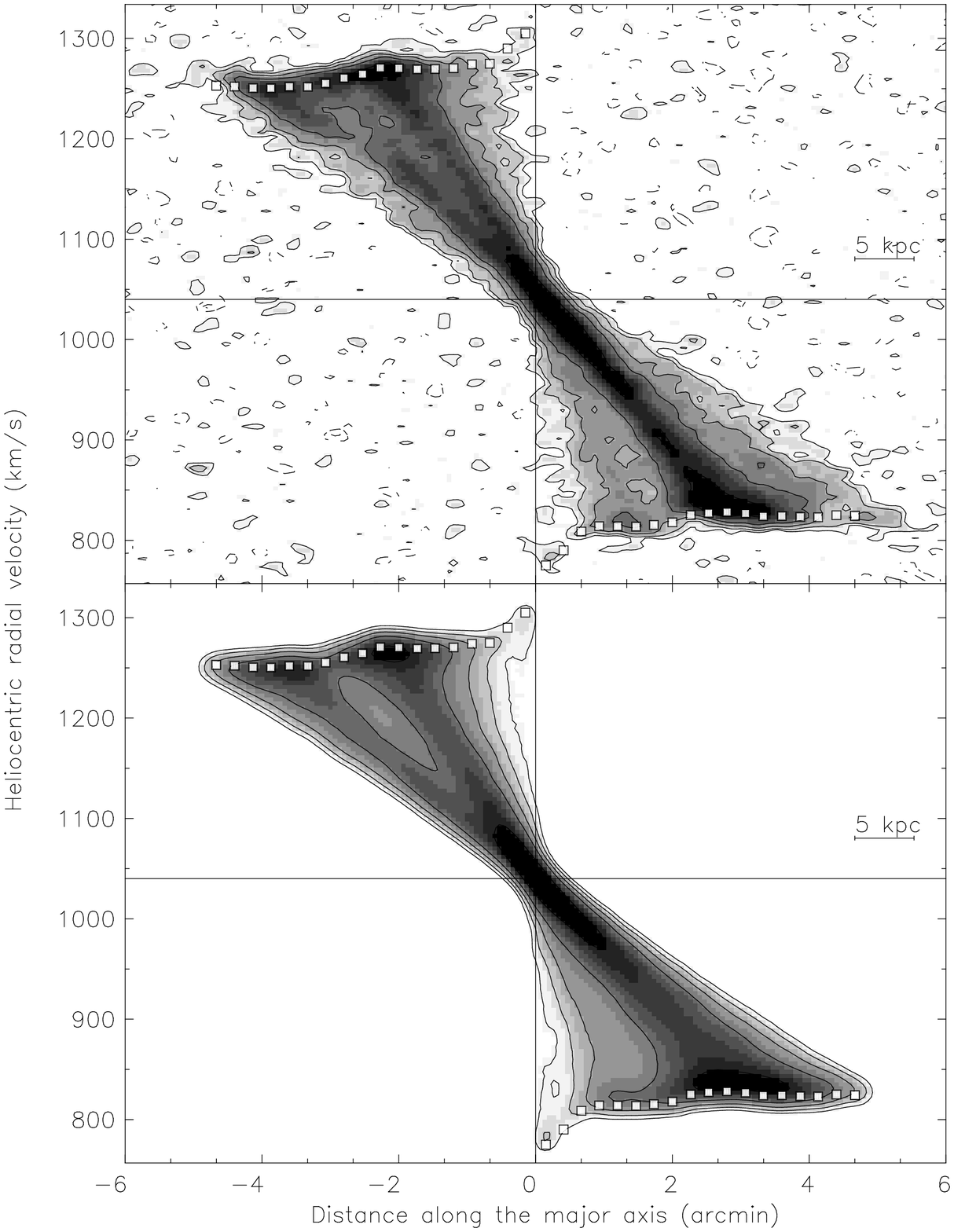}
 \caption{$Upper$: Position-velocity diagram along the major axis of the edge-on galaxy NGC\,7814. The squares show the rotation curves derived separately for the approaching and the receding side. The contour levels are: 0.5, 1, 2, 4, 10, 20 ${\rm mJy}\,{\rm beam}^{-1}$.
$Lower$: Position-velocity diagram along the major axis for the model galaxy. 
1 arcmin$=4.25 \kpc$.}
 \label{pv7814}
\end{figure}

\begin{figure*}
 \centering
 \includegraphics[width=\textwidth]{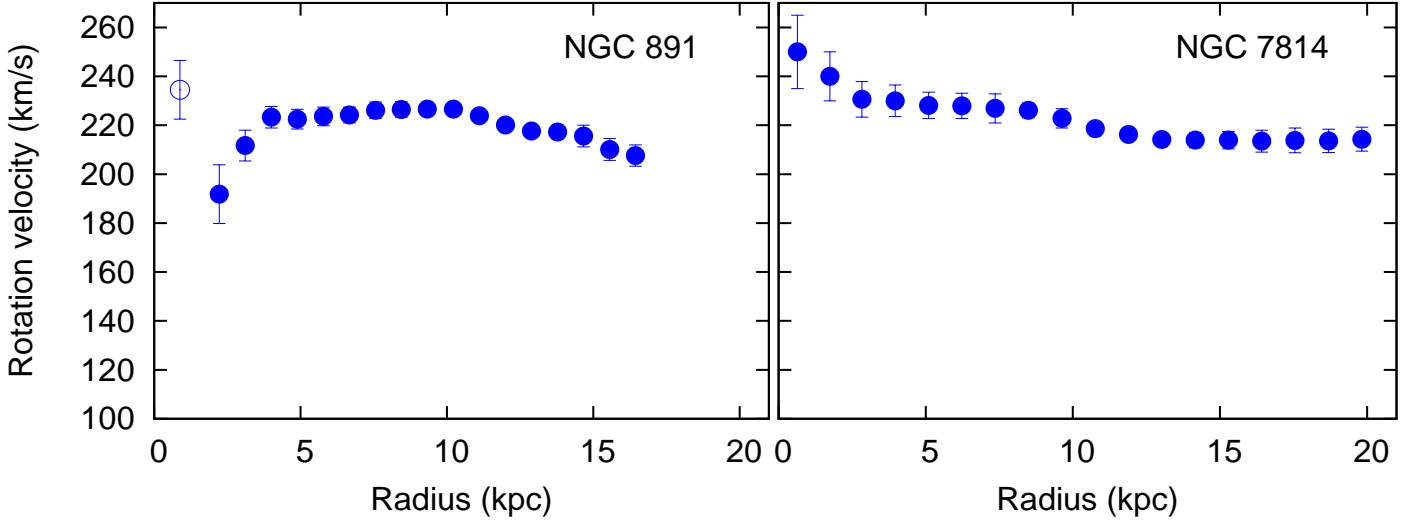}
 \caption{The rotation curves of NGC\,891 and NGC\,7814.
The open symbol for NGC\,891 outlines some potential contribution
from non-circular motions in the centre.
}
 \label{rotcurves}
\end{figure*}

The gas density as a function of $R$ was obtained by deprojecting the total \hi\ maps.
The model position-velocity diagram was compared (by eye) with the observed one.
When necessary, the initial input values of the rotational velocity were changed and the procedure was iterated until a satisfactory matching of model and observations was reached. 
This was done independently for the two sides of the galaxies. 
The final adopted rotation curves are shown plotted on the position-velocity maps (Figs.\ \ref{pv891} \& \ref{pv7814}). 
The averaged curves are shown in Fig.\ \ref{rotcurves}.
For each point, the error is the larger between the formal error of the fit and an uncertainty due to asymmetries between the approaching and the receding sides. 
The latter was calculated assuming that the difference between the two sides corresponds to a $2 \sigma$ deviation \citep{swaters99}.

The \hi\ disk of NGC\,891 is known to be lopsided \citep{baldwin80}.
On its southern receding side, it has an extended tail (Fig.\ \ref{pv891}) 
at a lower radial velocity with respect to the central part of the disk.
However, this decrease in radial velocity does not necessarily mean a decline in rotational velocity. 
As pointed out by \citet{sancisi79} it could be due to the location of the gas away from the line of nodes. 
Also, it is possible that non-circular motions dominate in these outer parts where the galaxy becomes so asymmetric. 
We derived, therefore, the rotation curve only for the symmetric part, inside $\sim6$ arcmin ($\sim 17 \kpc$) from the centre (Table \ref{tRC891}).
The sharp rise and peak near the centre ($R \lsim 1 \kpc$) indicate the presence of a fast rotating inner \hi\ disk or ring. 
Although this seems the straightforward explanation the possibility of a bar and associated non-circular motions cannot be ruled out \citep{garcia95}.
However, in the present data there is no indication of non-circular motions.
Just outside this central disk or ring the rotation curve is not well defined due to some beam-smearing and blending with the emission from the inner disk. 
The second point shown here is, therefore, rather uncertain.
We used different density distributions for the inner ring and estimate an error of about $12 \kms$.
In these central parts, the rotation curve is slightly asymmetrical: on the receding side it is somewhat steeper.
Further out the approaching and receding rotational velocities are the same within $<10 \kms$.
The rotation curve derived here differs from that obtained by 
\citet{sancisi79}
especially in the inner parts where the higher
sensitivity observations by \citet{Oosterloo+07} used here show the
inner steep rise and the fast-rotating component.  
The radial extent is about the same.

\begin{table}
\caption{Rotation curve of NGC\,891}
\label{tRC891} 
\centering    
\begin{tabular}{lcc}
\hline\hline
Radius & Radius & v$_{\rm c}$ \\
(arcmin) & (kpc) & ($\kms$)  \\
\hline
0.32  &  0.88   & $234.5 \pm 12.0$ \\
0.80  &  2.22   & $191.8 \pm 12.0$ \\
1.13  &  3.11   & $211.7 \pm 6.3 $\\
1.45  &  4.00   & $223.3 \pm 4.4$\\
1.77  &  4.89   & $222.5 \pm 4.0$\\
2.09  &  5.78   & $223.6 \pm 3.8 $\\
2.41  &  6.67   & $224.2 \pm 3.2$\\
2.73  &  7.56   & $226.1 \pm 3.4$\\
3.06  &  8.44   & $226.4 \pm 3.2$\\
3.38  &  9.33   & $226.6 \pm 2.7 $\\
3.70  &  10.22  & $226.6 \pm 2.7$\\
4.02  &  11.11  & $223.9 \pm 2.4$\\
4.34  &  12.00  & $220.1 \pm 2.4$\\
4.66  &  12.89  & $217.6 \pm 2.3  $\\
4.99  &  13.78  & $217.2 \pm 2.8$\\
5.31  &  14.67  & $215.6 \pm 4.4$\\
5.63  &  15.56  & $210.1 \pm 4.5$\\
5.95  &  16.44  & $207.6 \pm 4.4$\\
\hline
\end{tabular}
\end{table}

The rotation curve of NGC\,7814 (Table \ref{tRC7814}) has a very steep
rise near the centre followed by a slow decline which becomes a little
more pronounced beyond 2 arcmin ($\sim 8 \kpc$).
The curve in the inner 3 kpc from the centre is not accurately determined because of the poor signal/noise ratio. 
We estimate an error of about $15 \kms$.
The value of the rotational velocity near the centre (first two points) could 
be higher than given here. 
The third point of the rotation curve is well determined (see Fig.\ \ref{pv7814}).
There are no large asymmetries; only at large distances from the centre the two sides of the galaxy are not completely symmetrical. 
Fig.\ \ref{pv7814} shows that there are differences in the \hi\ density distribution and also in the kinematics. 
On the approaching side the rotational velocity begins to drop off around 2 arcmin from the centre whereas on the receding side the decrease seems to start a little further out.
This rotation curve is different from the flat curve derived by Van der Kruit and Searle (1982). 
Because of the better signal/noise ratio of the new observations it was possible to  derive it also in the inner region (within 1 arcmin $\sim 4 \kpc$ from the centre) where it has a steep rise to a maximum of $250 \kms$ at about $0.6 \kpc$ from the centre followed by a slow decline to about $214 \kms$ in the outer parts. 
Its radial extent is about the same.

\begin{table}
\caption{Rotation curve of NGC\,7814}
\label{tRC7814} 
\centering    
\begin{tabular}{lcc}
\hline\hline
Radius & Radius & v$_{\rm c}$ \\
(arcmin) & (kpc) & ($\kms$)  \\
\hline
0.15 &   0.64   & $250.0\pm  15.0  $\\
0.41 &   1.74   & $240.0\pm  10.0  $\\
0.67 &   2.83   & $230.6\pm  7.3  $\\
0.93 &   3.96   & $230.0\pm  6.5  $\\
1.20 &   5.10   & $228.1\pm  5.4  $\\
1.47 &   6.23   & $227.9\pm  5.2  $\\
1.73 &   7.36   & $226.9\pm  5.9  $\\
2.00 &   8.49   & $226.1\pm  3.0  $\\
2.27 &   9.63   & $222.8\pm  3.9  $\\
2.53 &   10.76  & $218.6\pm  2.9  $\\
2.80 &   11.89  & $216.2\pm  2.6  $\\
3.07 &   13.02  & $214.2\pm  2.6  $\\
3.33 &   14.16  & $213.9\pm  3.1  $\\
3.60 &   15.29  & $213.9\pm  3.6  $\\
3.87 &   16.42  & $213.5\pm  4.5  $\\
4.13 &   17.55  & $213.8\pm  5.1  $\\
4.40 &   18.69  & $213.6\pm  4.8  $\\
4.67 &   19.82  & $214.3\pm  4.9  $\\
\hline  
\end{tabular}
\end{table}

\section{Comparison of the rotation curves with the distribution of light}

The rotation curves for the two galaxies are compared in Fig.\ \ref{rotcurves}.
They have similar amplitudes but significantly different shapes.
The rotation curve of NGC\,7814 indicates a more pronounced mass concentration to the centre than in NGC\,891. 
The correspondence with the central concentration of light in NGC 7814 seems evident.
Given the potential uncertainties in the interpretation of the first point 
for NGC\,891, we indicate it as an open symbol.
In order to further investigate the distribution of the mass as compared to that of the light in the two galaxies we made the standard decomposition of the rotation curves in mass components: bulge, stellar and gaseous disks, and dark matter halo.

\subsection{Photometric data}

In order to avoid problems with dust extinction we used data in the $3.6\,\mu{\rm m}$ band obtained with the Spitzer Space Telescope \citep{spitzer}.
Both NGC\,891 and NGC\,7814 had been observed with the Spitzer Telescope and the mosaics were already available in the archive.
The observations of NGC\,891 had an exposure time of 96.8$\,s$, those of NGC\,7814 26.8$\,s$.
We converted from the Spitzer units of ${\rm MJy}\,{\rm sr}^{-1}$ to 
$\magasec$  in $3.6\,\mu{\rm m}$ band assuming an absolute magnitude for 
the Sun of $M_{3.6\,\mu{\rm m}}=3.24$ \citep{Oh+08}.
We applied the surface brightness correction simply by multiplying our fluxes by 0.91 (Spitzer Handbook).
We estimated the values of the background in the two images to be 21.06 and $20.70\magasec$ respectively for NGC\,891 and NGC\,7814. 
The $3.6\,\mu{\rm m}$ images for the two galaxies are shown in Fig. 5 and in Fig. 6.

\begin{figure*}[ht]
 \centering
 \includegraphics[width=0.45\textwidth, angle=-90]{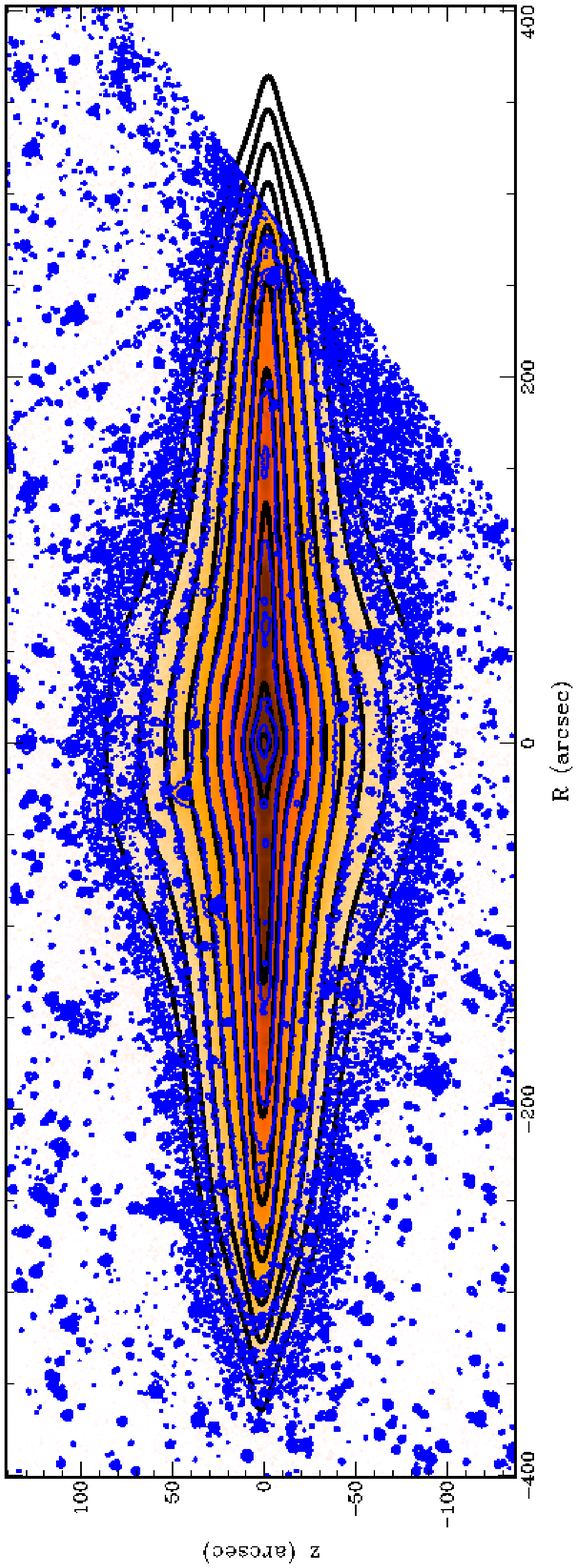}
 \includegraphics[width=0.495\textwidth]{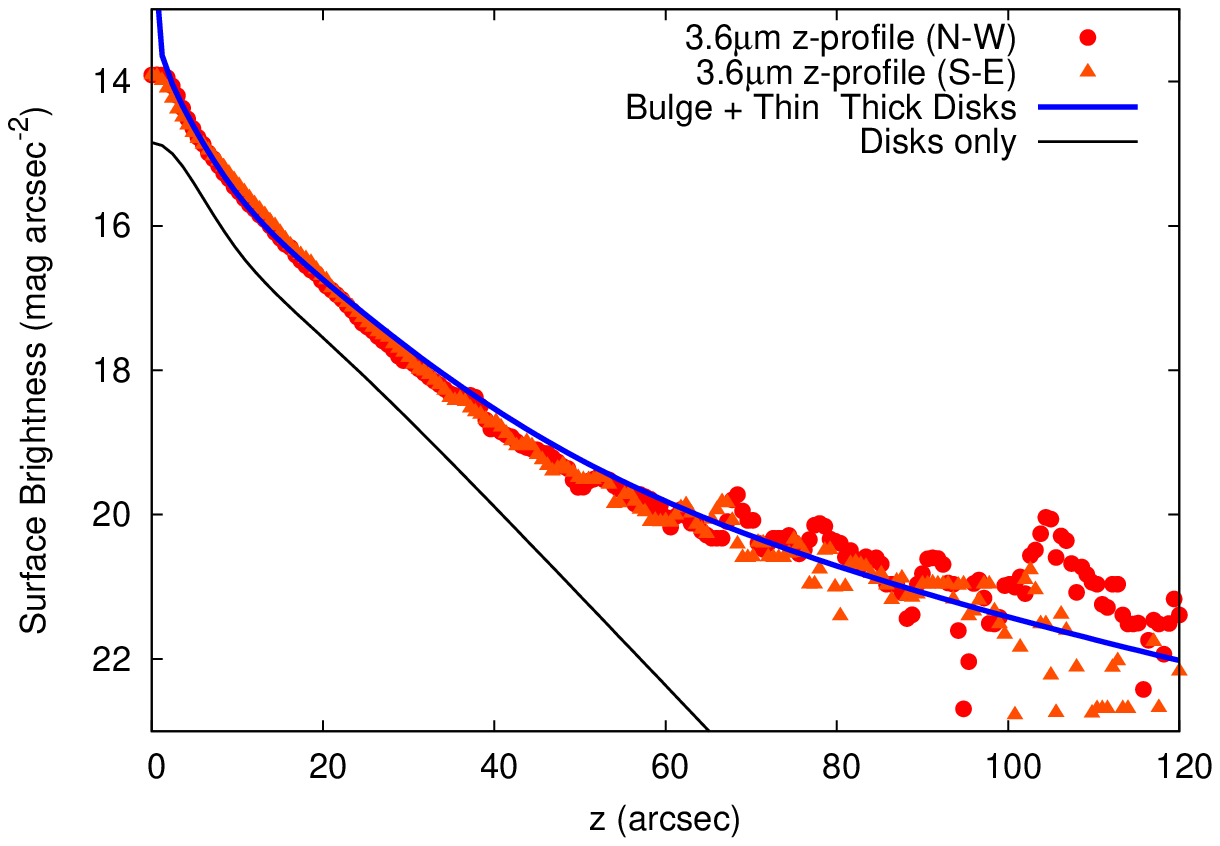}
 \includegraphics[width=0.495\textwidth]{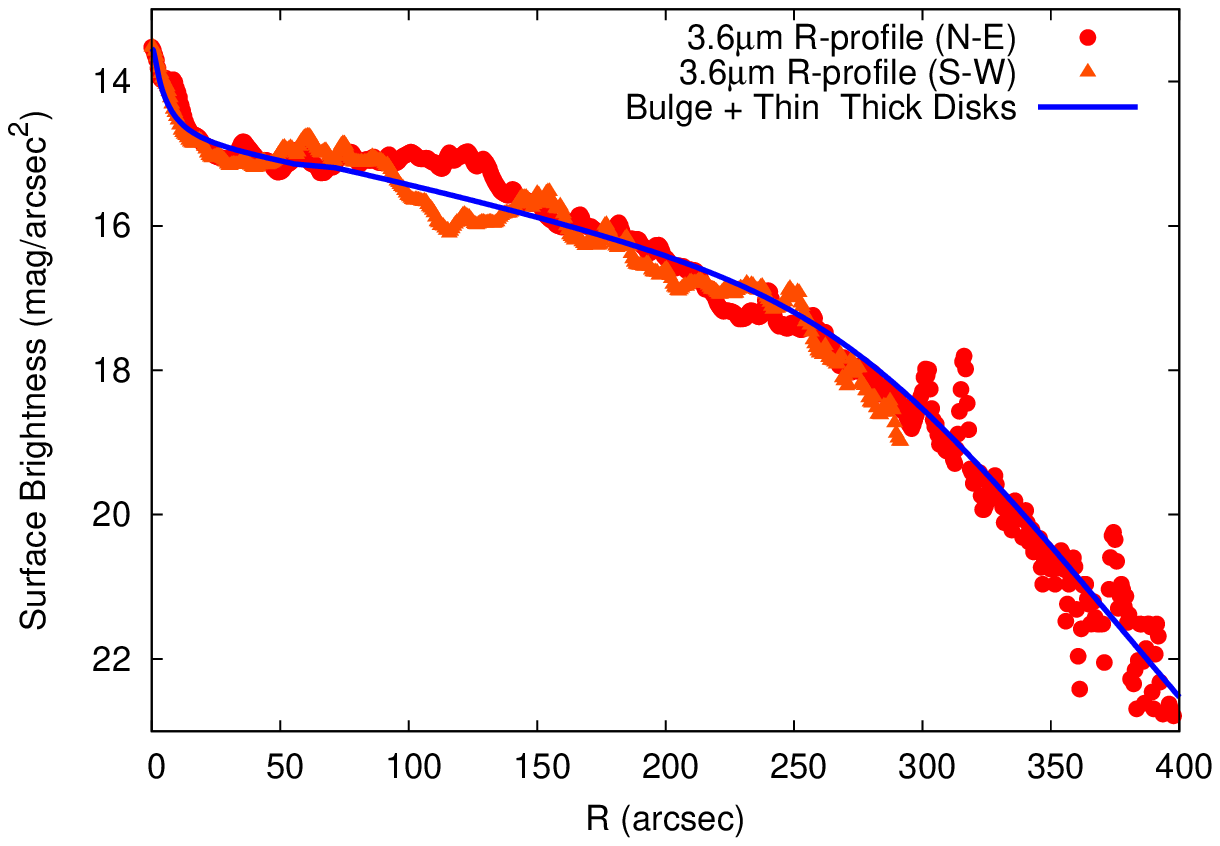}
 \caption{
 $3.6\,\mu{\rm m}$ photometry for NGC\,891. \emph{Top panel:} distribution of light 
in the sky in thin (blue) contours overlaid with our best model with bulge and
thin and thick disks, see Table \ref{tPhot}.
\emph{Bottom left:} Light distribution along the vertical direction
above and below the plane (circles and triangles) compared with our best model
(thick blue curve).
\emph{Bottom right:} Light distribution along the plane of the galaxy
(circles and triangles) compared with our best model (thick blue curve).
}
 \label{galfit891}
\end{figure*}

We performed a bulge-disk decomposition of the stellar light using GALFIT \citep{galfit1, galfit2}.
For both galaxies we extracted the psf directly from the image taking a relatively isolated star in the field close to the galaxies.
The images were masked in order to exclude stars, background galaxies and spurious features of the CCD.
We used  a S\'ersic profile for the bulge component and a perfectly edge-on disk for the stellar disk.
NGC\,891 could not be fitted with a single disk but it required both a thin and a thick disk.
The presence of two disks had been found also before in other bands \citep{vdkruit81, ShawGilmore89, Xilouris98}. 
We note that the functional form of the density distribution in the vertical direction has a strong impact on the properties of the disk(s) required by the fit.
At the moment GALFIT allows only for a ${sech}^2$ vertical density profile, but it is possible that with an exponential profile the two components could have very different parameters or the second disk might even not be required.

In fitting the bulge and the disk(s) we first fixed the centres and the position angles of the various components to the same value. 
For NGC\,7814, we fixed the S\'ersic index to 4, i.e.\ we assumed a de Vaucouleurs profile;
leaving the S\'ersic index free gives values very close to 4.
In NGC\,891 
there is a degeneracy between the scale-length of the bulge and the 
scale-height of the thick disk and, to obtain a satisfactory result, we 
preferred to fix the latter to a value of $h_{\rm z}=0.8 \kpc$.
Once these
values are fixed, the simultaneous fits of all the other 
parameters converge for both galaxies.

\begin{figure*}[ht]
 \centering
 \includegraphics[width=0.33\textwidth, angle=-90]{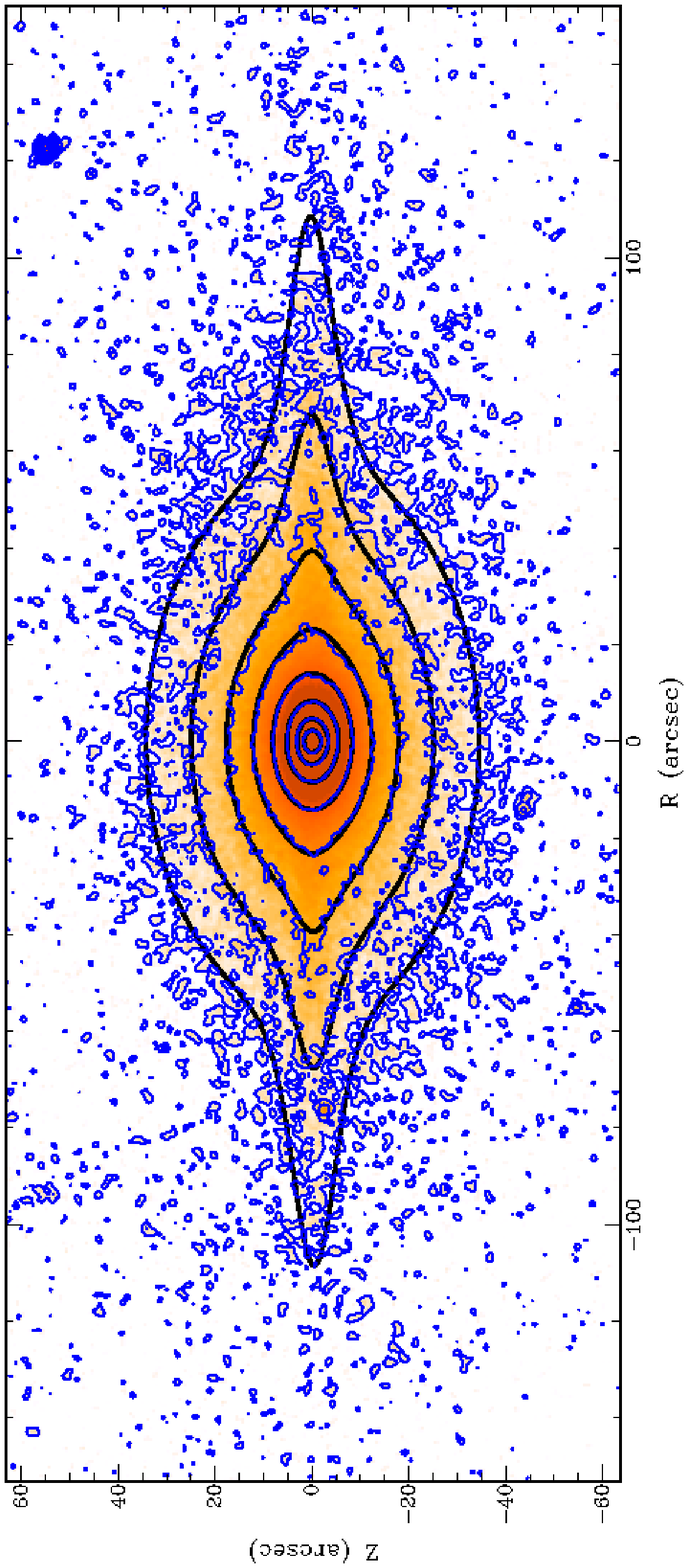}
 \includegraphics[width=0.495\textwidth]{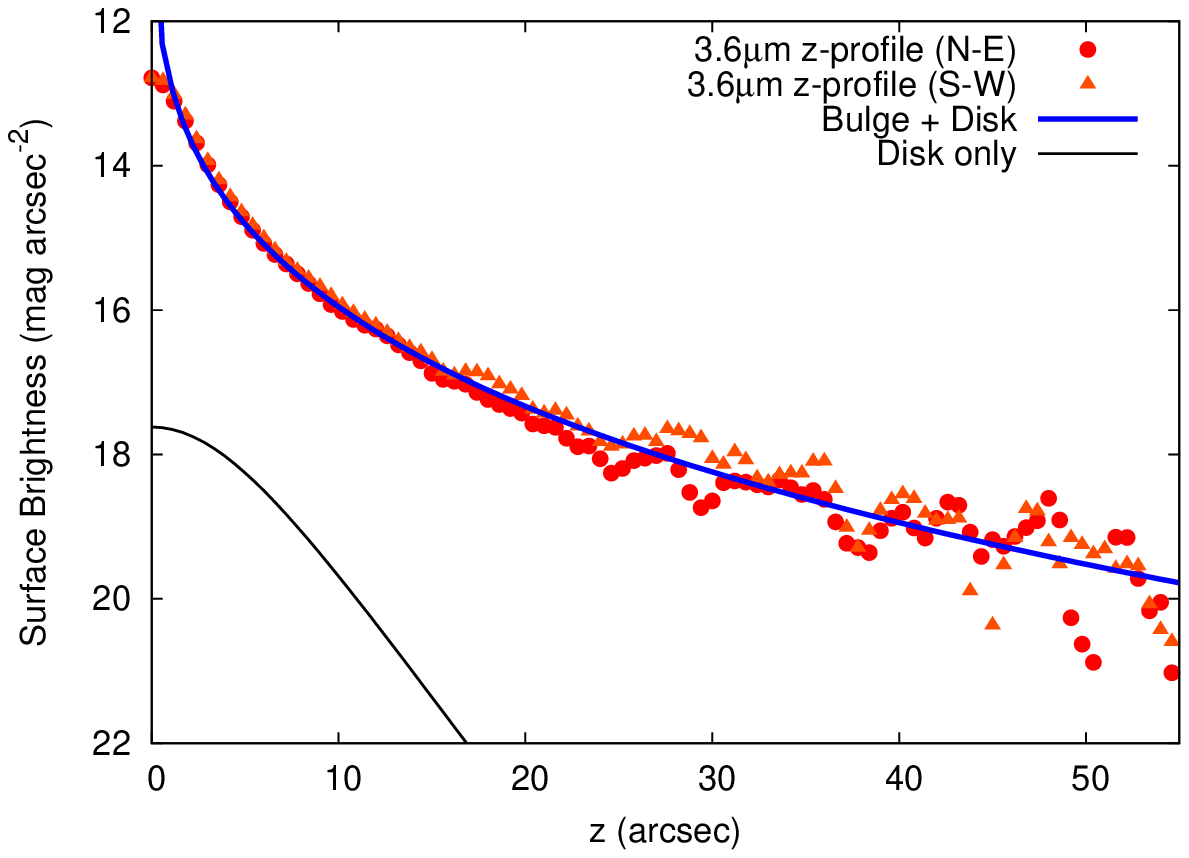}
 \includegraphics[width=0.495\textwidth]{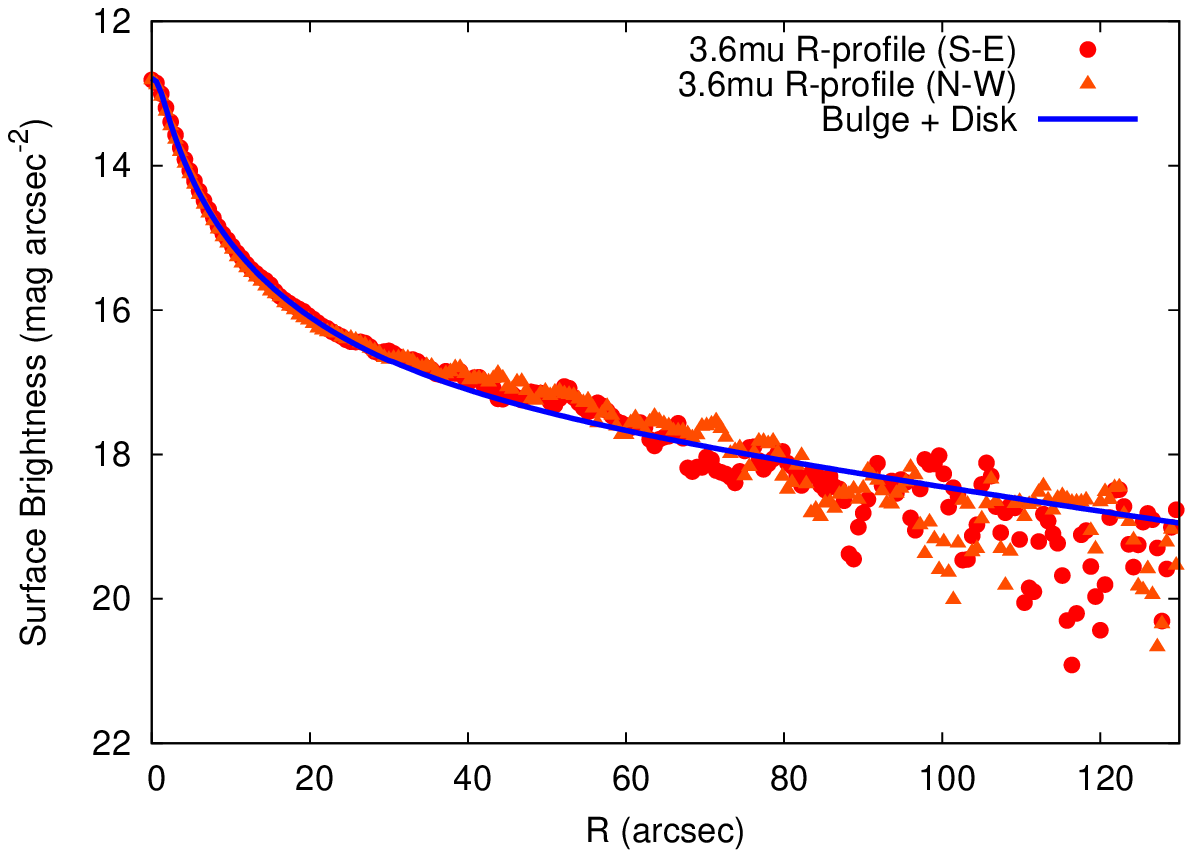}
 \caption{
$3.6\,\mu{\rm m}$ photometry for NGC\,7814. \emph{Top panel:} distribution of light 
in the sky in thin (blue) contours overlaid with our best model with bulge and
disk, see Table \ref{tPhot}.
\emph{Bottom left:} Light distribution along the vertical direction
above and below the plane (circles and triangles) compared with our best model
 (thick blue curve).
\emph{Bottom right:} Light distribution along the plane of the galaxy
(circles and triangles) compared with our best model (thick blue curve).
}
 \label{galfit7814}
\end{figure*}

Figs.\ \ref{galfit891} and \ref{galfit7814} show our best photometric models
for NGC\,891 and NGC\,7814 respectively.
In both figures the top panels show the light distribution (at $3.6\, \mu{\rm m}$)
in color shade and thin (blue) contours from the Spitzer mosaic.
The thick (black) contours show the GALFIT solutions obtained
using two disks $+$ a S\'ersic (n=3) bulge for NGC\,891 and one disk $+$ a
$r^{-1/4}$ bulge for NGC\,7814 (see Table \ref{tPhot}).
The bottom panels in Figs.\ \ref{galfit891} and \ref{galfit7814} show 1D cuts
of the data and the models along the vertical directions (left) and
along the plane of the galaxies (right).
In the plots perpendicular to the plane we show the data from both
sides (circles and triangles), 
our best model (thick blue curve) and the contribution of the disk only 
(thin black curve).
Note that this latter is completely negligible for NGC\,7814.
In the plots along the plane for NGC\,891 one can appreciate that the disk
in not exponential.
This shape has been modelled using the truncation function implemented in
GALFIT with $r_{\rm break}=120''$ and $\Delta\,r_{\rm soft}=280''$
\citep[see][]{galfit2}.
The truncation is the same for both the thin and the thick disk.
The disk of NGC\,7814 does not require any truncation.

\begin{table}
\caption{Spitzer $3.6\,\mu{\rm m}-$band photometric parameters for the two galaxies}  
\label{tPhot} 
\centering    
\begin{tabular}{lccc}
\hline\hline             
Parameter & \multicolumn{2}{c}{NGC\,891} & NGC\,7814 \\ 
\hline
& Thin disk & Thick disk & \\
$h_{\rm R}$ (kpc)         &   4.18 & 5.13              &     4.26           \\
$\Sigma_e$ ($\loSpc$)    &   443.5 & 368.9              &     78.0            \\
$h_{\rm z}$ (kpc)         &   0.25 & $0.80^a$              &     0.44           \\
$L_{\rm disk}$ ($\loS$)   & $4.4 \times 10^{10}\,^b $&$5.1 \times 10^{10}\,^b $  & $8.4\times 10^{9}$ \\
\hline
S\'ersic index            &  \multicolumn{2}{c}{2.99}   & 4.0$^a$ \\
$r_{\rm e}$ (kpc)         & \multicolumn{2}{c}{1.80} &    2.16             \\
$I_{\rm e}$ ($\loSpc$)    & \multicolumn{2}{c}{525.8}& $1.12\times 10^{3}$ \\
$q$                      & \multicolumn{2}{c}{0.68} &   0.61             \\
$L_{\rm bulge}$ ($\loS$)  & \multicolumn{2}{c}{$2.2\times 10^{10}$} & $7.0\times 10^{10}$ \\
\hline
\end{tabular}\\
\begin{flushleft}
$^a$ fixed in the fit; $^b$ calculated out to $R=17\kpc$ without considering the truncation.
\end{flushleft}
\end{table}

Table \ref{tPhot} gives the photometric parameters of the fits shown in 
Figs.\ \ref{galfit891} and \ref{galfit7814}.
Since our rotation curve fitting routine requires a spherical bulge we used a spherical-equivalent effective radius $r_{\rm e, sph} = r_{\rm e} \sqrt{q}$ (geometrical mean) where $q$ is the axis ratio.
The $3.6\,\mu{\rm m}-$band disk and bulge luminosities in Table \ref{tPhot} are calculated 
out to the last measured points of the rotation curves, $R_{\rm max} = 17 \kpc$ and $20 \kpc$ 
for NGC\,891 and NGC\,7814 respectively.
The values derived for the NGC\,891 disks do not take into account the truncations and are therefore upper limits.
As expected, the dominant luminosity component in NGC\,891 is that of the disks, about four times brighter than the bulge.
On the contrary, in NGC\,7814 the bulge is totally dominating with a 
bulge-to-disk ratio in the $3.6\,\mu{\rm m}-$band of about 9.
In K-band this ratio is about 14 \citep{wainscoat90}.

\subsection{Maximum light (bulge + disk)}

\begin{figure*}[ht]
 \centering
 \includegraphics[width=0.49\textwidth]{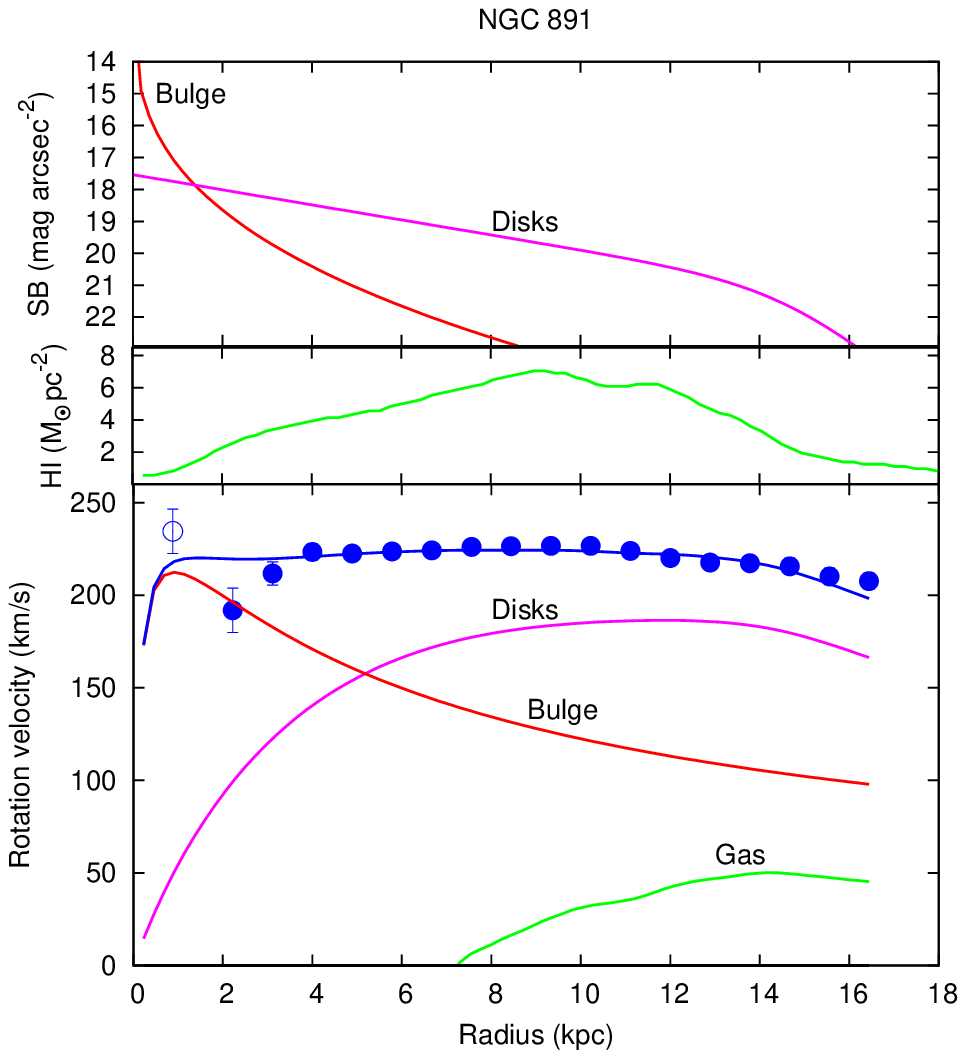}
 \includegraphics[width=0.49\textwidth]{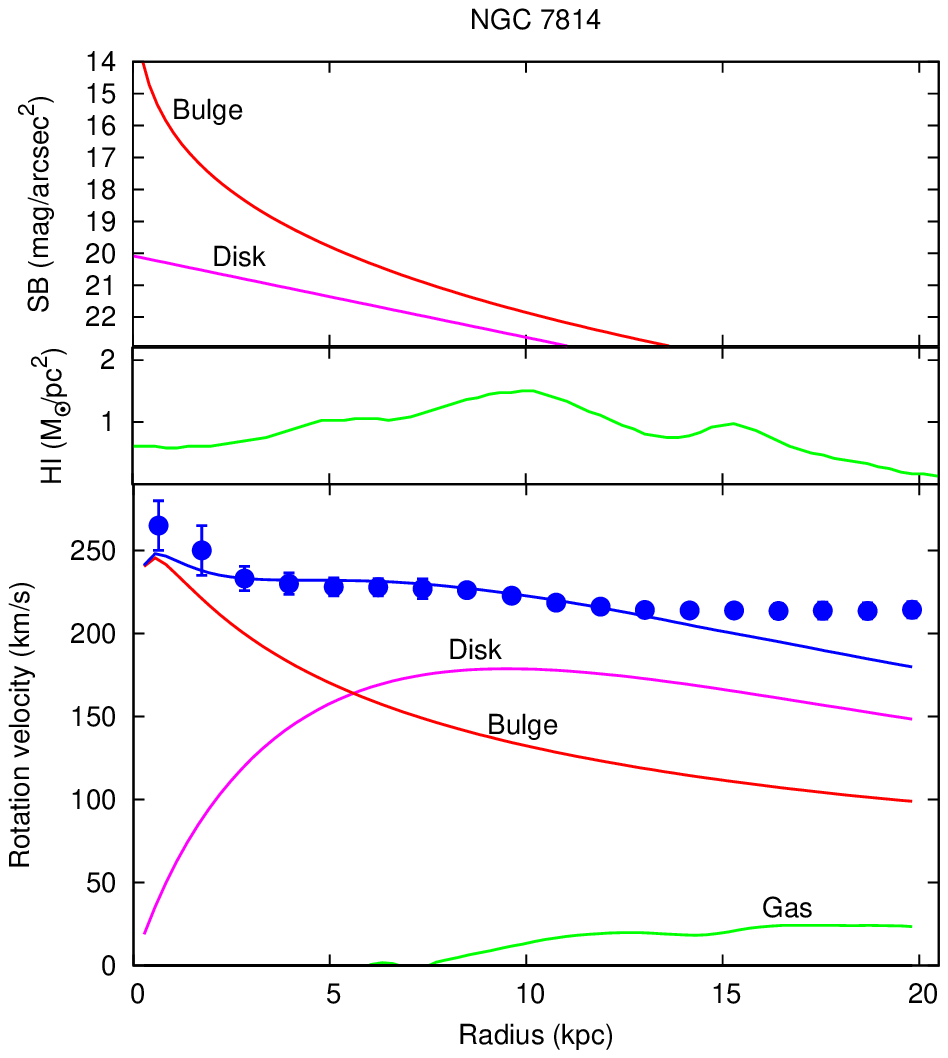}
 \caption{Rotation curve decompositions for NGC\,891 and NGC\,7814. 
\emph{Top panels:} surface brightness profiles at $3.6\, \mu{\rm m}$ for the bulge and
the stellar disks built from the parameters of Table \ref{tPhot}
(for NGC\,891 the two disks have been added). 
\emph{Middle panels:} \hi\ surface density.
\emph{Bottom panels:} Best fit without a dark matter halo, the $\ml$ ratios
of bulge and disk components are given in Table \ref{tDecomp}.
$1\kpc=21.7''$ and $33.4''$ for NGC\,891 and NGC\,7814 respectively.
}
 \label{decompositions_nohalo}
\end{figure*}

Fig.\ \ref{decompositions_nohalo} (upper panels) shows the $3.6\,\mu{\rm m}-$band 
photometric profiles adopted for bulge and disk. 
For NGC\,891 the thin and thick disks are combined in a single component with intermediate scale-height.
The truncation in GALFIT is applied to the projected edge-on profile and it
requires deprojection to a face-on view.
If the parameter $\Delta\,r_{\rm soft}$ is small compared to $r_{\rm break}$ 
this can be achieved with good accuracy by simply applying the truncation
function to the face-on profile (C.\ Peng, priv. comm.).
However in our case $\Delta\,r_{\rm soft} > r_{\rm break}$ and the deprojection
is more complex.
In order to find a functional form for the face-on truncation 
we built several mock edge-on truncated disks with GALFIT and
deproject the light profiles using the ``Lucy-method'' 
\citep{Warmels88}.
We then fitted these profiles with exponential disks multiplied by the
GALFIT truncation function.
We found that good results are obtain using for the face-on view the same
$\Delta\,r_{\rm soft}$ as for the edge-on view but a $R_{\rm break (face-on)}
\approx r_{\rm break} + \frac{1}{e} \Delta\,r_{\rm soft}$.
The effect on the light profile is shown in the upper left panel of 
Fig.\ \ref{decompositions_nohalo}.

The bottom panels of Fig.\ \ref{decompositions_nohalo} show the rotation 
curve decompositions for NGC\,891 and NGC\,7814 using the above 
$3.6\,\mu{\rm m}-$band photometry. 
We show here the maximum light (bulge + disk) solutions for both galaxies. 
The fit was obtained by 
fitting simultaneously the $\mlS$ ratios of bulge and disks 
to the points inside about 3 scale-lengths ($\sim 15 \kpc$ for NGC\,891
and $\sim 13 \kpc$ for NGC\,7814) as it is customary for maximum light
tests \citep{vanAlbada85}.
The gas surface density (middle panels)
was multiplied by a factor 1.4 to account for Helium.
The fitting parameters are given in Table \ref{tDecomp}.

In the fit for NGC\,891 the bulge dominates in the inner parts and
accounts for the steep rise and the inner peak of the rotation curve. 
The first point of the rotation curve was  included in this fit in
spite of the uncertainties about its nature and origin 
pointed out above. 
However, the bulge $\ml$ and the quality of the fit
would not change significantly, first point included or not.
At larger radii, beyond $\sim 5 \kpc$, the disk dominates. 
Its contribution to the total mass is about three times as large as
that of the bulge.  
The shape of the rotation curve is remarkably well reproduced out to
the last point. There is no discrepancy between observed and model
curve and no dark matter halo is required here. 
This, however, is not too surprising. The rotation curve
is not very extended: it is not traced beyond the bright optical disk
and out to radii where usually the halo becomes conspicuous. 
The values obtained for the $\ml$ ratios of bulge and disk (Table
\ref{tDecomp}) seem reasonable \citep[cf.][]{verheijen97}. 

\begin{table}
\caption{Fits to the rotation curves}
\label{tDecomp} 
\centering    
\begin{tabular}{lcc}
\hline\hline
Parameter                            & NGC\,891           & NGC\,7814 \\ 
\hline
~~Maximum light &  &\\
\hline
Bulge $\mlS$                         & $1.64\pm 0.07$ & $0.64 \pm0.03$ \\
Disk $\mlS$                          & $0.90 \pm 0.02$ & $9.25 \pm0.27$ \\
Reduced $\chi^2$                     & $1.05^a$         & $0.59^a$ \\
\hline
~~ Isothermal halo & &\\
\hline
Bulge $\mlS$                         & $1.63 \pm 0.25$      & $0.71\pm0.05$\\ 
Disk $\mlS$                          & $0.77 \pm 0.16$      & $0.68 \pm 0.94$ \\
DM halo $\rho_{\rm 0}$($10^{-3}\mo\,{\rm pc}^{-3}$) & $33.1\pm 16.0$ & $152.4\pm 95.7$ \\ 
DM core radius $r_{\rm 0}$ (kpc)     & $1.9 \pm 4.4$         & $2.1 \pm 0.6$\\ 
Reduced $\chi^2$                    & 1.30                  & 0.39 \\
\hline
\end{tabular}\\
$^a$Referred to the radial range where the fit is performed (within 3 scale-lengths), see text.
\end{table}

\begin{figure*}
 \centering
 \includegraphics[width=0.49\textwidth]{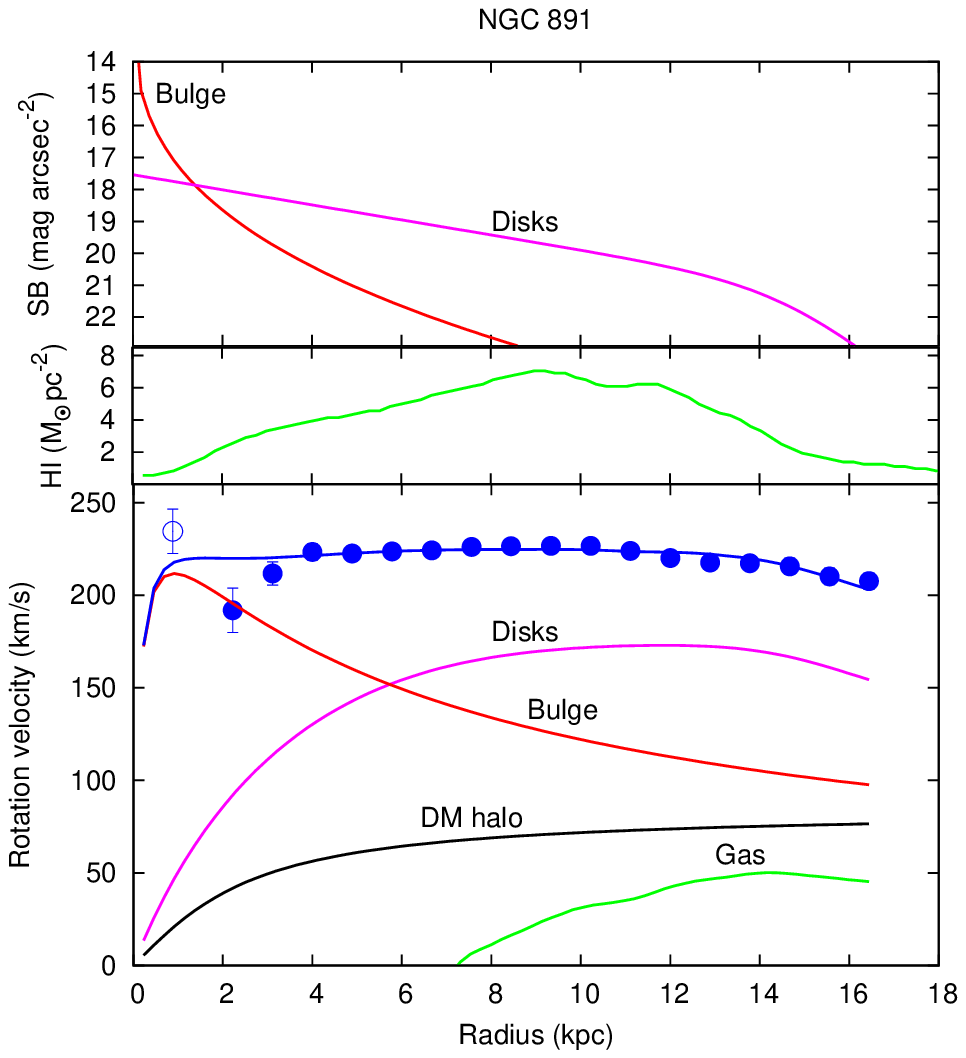}
 \includegraphics[width=0.49\textwidth]{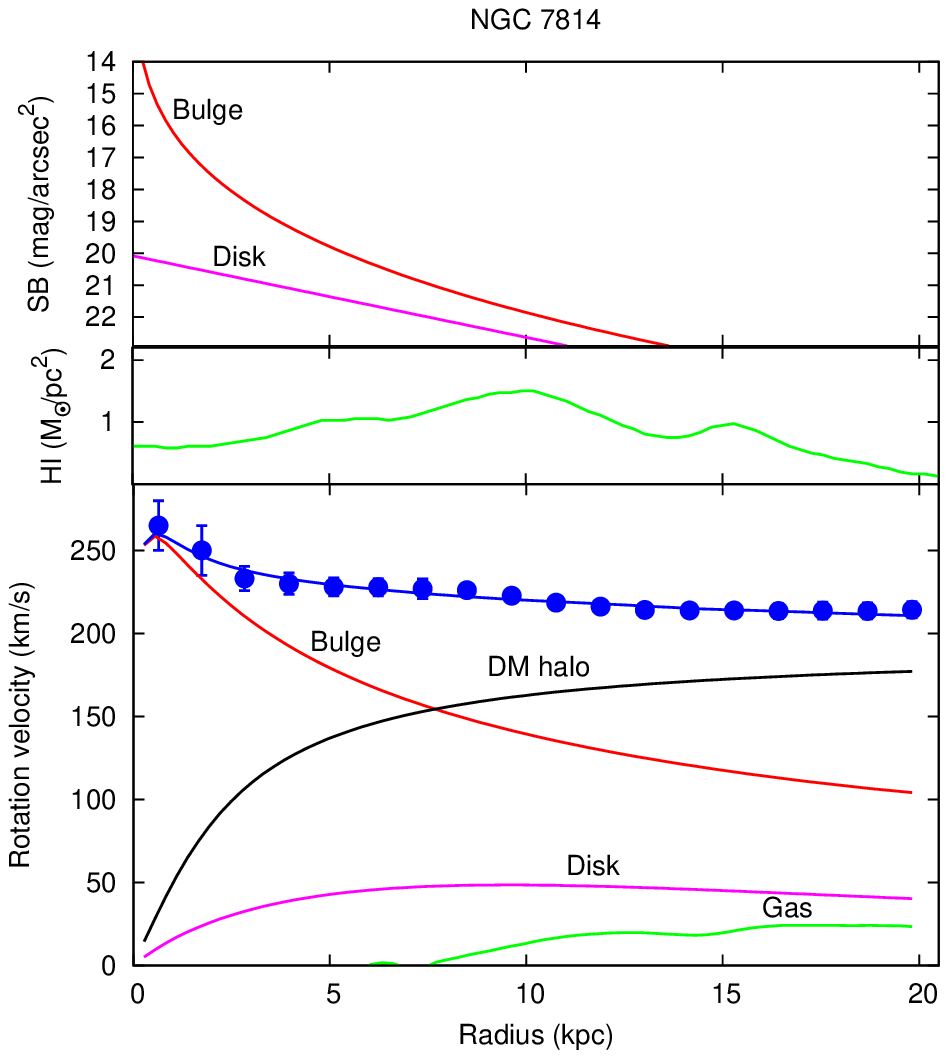}
 \caption{Rotation curve decompositions for NGC\,891 and NGC\,7814. 
\emph{Top panels:} surface brightness profiles at $3.6\, \mu{\rm m}$ for the bulge and
the stellar disks built from the parameters of Table \ref{tPhot}
(for NGC\,891 the two disks have been added). 
\emph{Middle panels:} \hi\ surface density.
\emph{Bottom panels:} Best fit with an isothermal dark matter halo, 
the parameters of the fits are given in Table \ref{tDecomp}.
$1\kpc=21.7''$ and $33.4''$ for NGC\,891 and NGC\,7814 respectively.
}
 \label{decompositions_free}
\end{figure*}

In conclusion, for NGC 891 a maximum light (bulge+disk) solution is quite satisfactory.
Bulge and disk together are able to explain the rotation curve. 
Their relative contributions, however, are uncertain given the mentioned
degeneracy between bulge and thick disk.
It is clear, at any rate, that in NGC 891 the disk, which is responsible for most of the light, contributes also the largest part of the mass.

In the NGC\,7814 maximum-light decomposition the bulge dominates in
the inner parts.  
Bulge and disk together provide an excellent fit to the rotation curve
out to $R \sim 13 \kpc$.
Beyond that, a mild discrepancy begins to show up between observed and
model curve.  
This is the kind of discrepancy found in the outer parts of spiral
galaxies and usually interpreted as evidence for the presence of a
dark matter halo. 
The $\mlS$ for the bulge is close to normal, whereas that for the disk
is unrealistically high. 
An alternative to such a heavy ``dark'' disk would be a substantial
contribution from a dark matter halo (see below). 
The presence of a massive bulge component in this galaxy is dictated
by the shape of the rotation curve in the inner parts. 
The bulge is completely dominant in the inner $5 \kpc$. 
In conclusion, a maximum light solution for NGC\,7814 provides an
excellent fit for most of  the rotation curve (out to $13 \kpc$) but 
it requires an unrealistically high $\mlS$ for the disk. 
A dark matter (DM) component is, therefore, needed for the region of the
disk and to account for the discrepancy in the very outer parts. 
Clearly, there is for NGC\,7814 a disk/halo degeneracy.

\subsection{Fits with isothermal DM halos}

The maximum light solutions investigated above are useful to
understand the possible connection with the light distribution and the
role played by baryons. 
They show that the distribution of light and that of mass are very
similar inside each of the two galaxies.
Here we add dark matter halos to the fits and
model the DM distribution as a standard isothermal sphere
\citep{vanAlbada85}.
We fit the four parameters ($\ml$ ratios and halo parameters)
simultaneously. 
The resulting fits are shown in Fig. \ref{decompositions_free} and
the values of the parameters are given in Table \ref{tDecomp}.
Using a NFW profile \citep{NFW97} would not change significantly
our results.
The best-fit values of the concentration parameters are above 10 for both 
galaxies.
If we fix them to $c=10$ we obtain $\ml$ ratios comparable to
those in Table \ref{tDecomp}, except for the $\ml$ of the disk in the 
free fit of NGC\,7814, which becomes $\mlS=2.7$; if however we fix this
latter to 0.7 the fit is still acceptable.

In NGC 891 the results are very similar to those obtained above with
the maximum light solution. 
The fit is still excellent. 
The values for the $\mlS$ of bulge and disk differ only slightly and
the dark halo plays a minor role. 
The exclusion of the first point of the rotation curve in the fit 
would not make any difference.

NGC\,7814 is more puzzling than NGC\,891 because of the disk/halo
degeneracy.
However, a fit with all four parameters ($\ml$ ratios and halo parameters)
free converges and gives acceptable results.
The bulge is still dominant in the inner $7-8 \kpc$ and very close to
maximum with only a minimal change in $\mlS$.
Further out, the halo now completely dominates in the place of the
disk.

These results clearly indicate that the two main luminous components in the 
two galaxies, the disk in NGC 891 and the bulge in NGC 7814, are closely 
linked to the distributions of mass as traced by the rotation curves.

\subsection{MOND}

\begin{figure}
 \centering
 \includegraphics[width=0.49\textwidth]{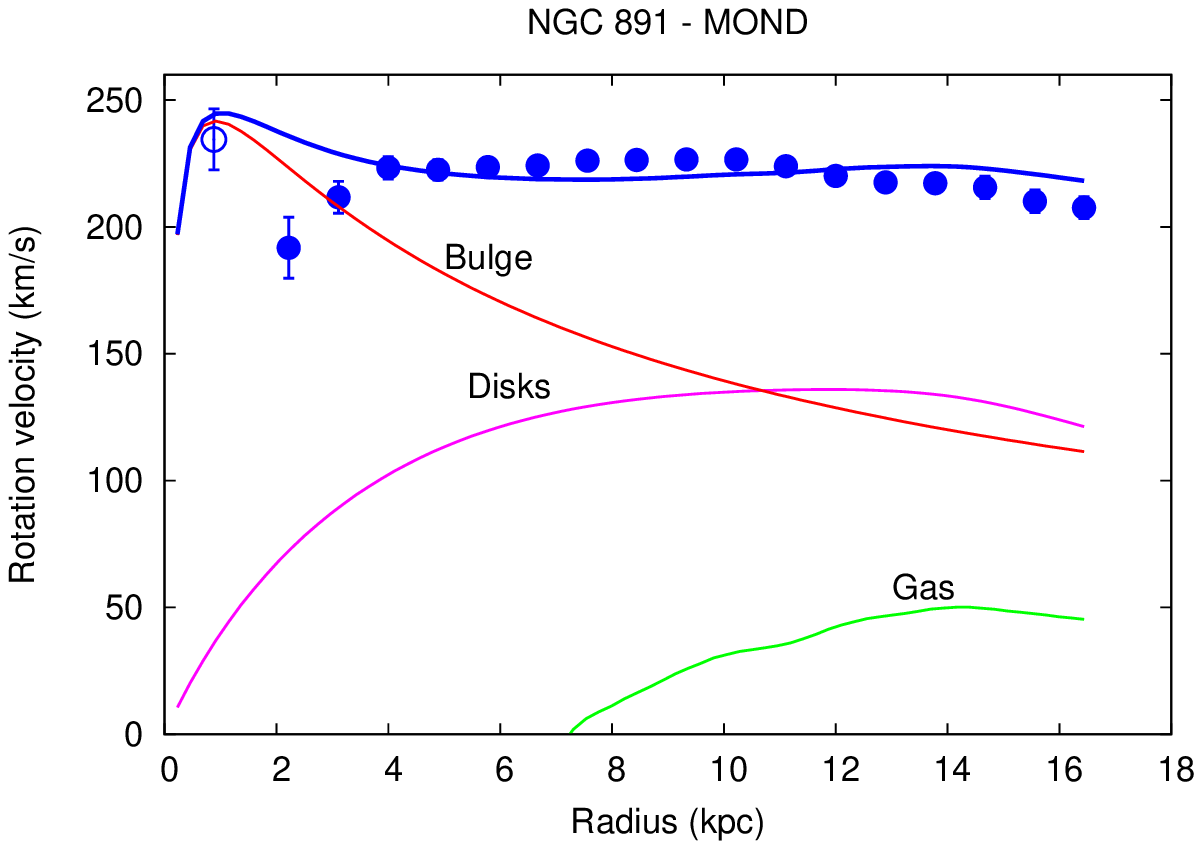}
 \includegraphics[width=0.49\textwidth]{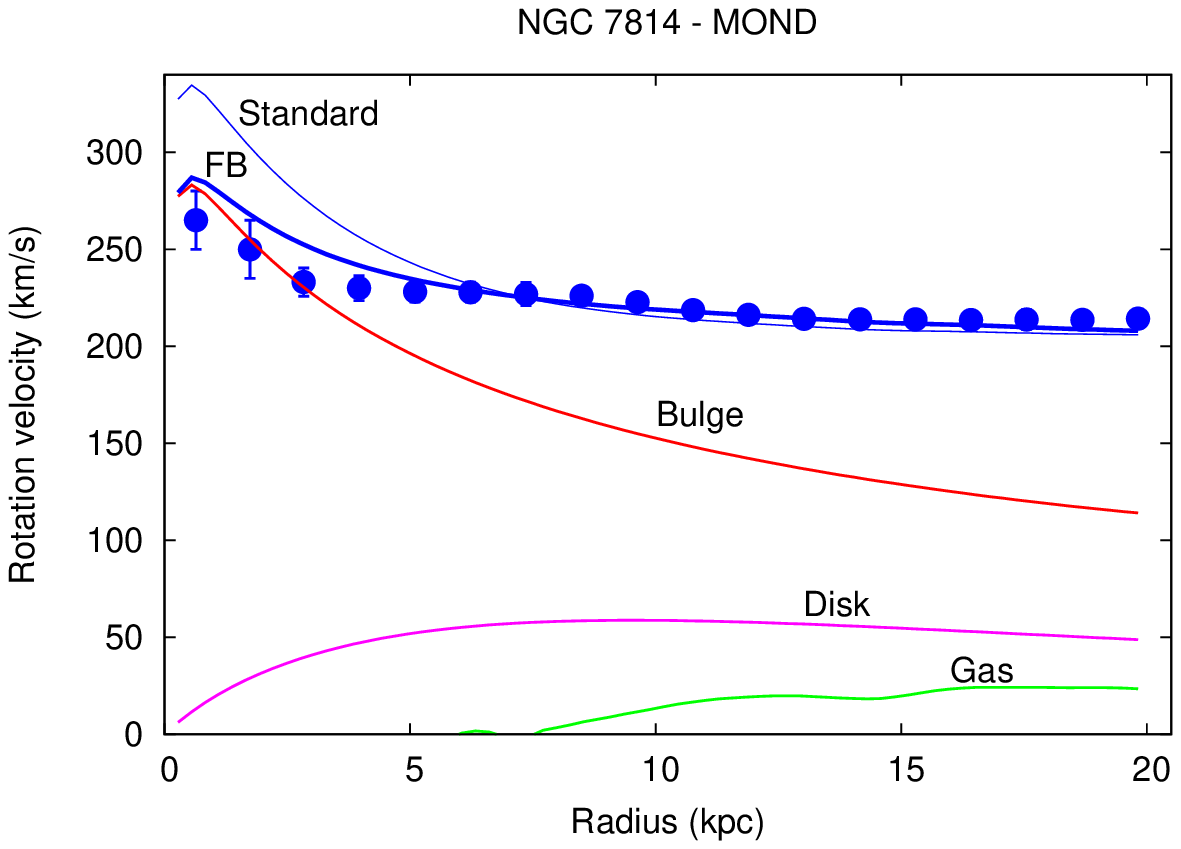}
 \caption{Comparison between the rotation curves of NGC\,891 and
   NGC\,7814 and the MOND predictions.
For NGC\,7814 we show the predictions obtained using the standard
interpolation function (thin blue curve) and the so-called ``simple'' 
(FB) function of \citet{famaey05} (thick blue curve).
}
 \label{MOND}
\end{figure}

We also compared both rotation curves with the predictions from the
MOdified Newton Dynamics (MOND) \citep{Milgrom83}. 
The rotation curve of NGC\,891 is reproduced, using the standard value
of $a_{\rm 0}=3700\,{\rm km}^2 {\rm s}^{-2} {\rm kpc}^{-1}$ with
best-fit $\ml$ ratios for the bulge and disk components of 2.0 and 0.5
respectively.
The result is shown in the upper panel of Fig.\ \ref{MOND}.
The quality of this fit is however not very good leading to a reduced $\chi^2$
of 4.8.

For NGC\,7814 a MOND fit with free $\mlS$ ratios for the bulge and disk
components requires for the latter a very high value of 4.6.
This is needed to fit the outer parts of the rotation curve.
In this case the $\mlS$ of the bulge is 0.77.
If the $\mlS$ of the disk is kept fixed to a more acceptable value of
1, the MOND fit using the standard interpolation function of 
\citet{Milgrom83} is not acceptable ($\chi^2=7.2$).
The bottom panel of Fig.\ \ref{MOND} shows the model predictions for
this function (thin curve) and also for the so-called ``simple'' function
of \citet{famaey05}.
Clearly this second function gives a much better fit ($\chi^2=1.3$
bulge $\mlS=0.83$).
We also used the ``simple'' function for NGC\,891 but we did not find
a significant improvement of the fit (disk $\mlS=0.23$, bulge
  $\mlS=1.9$, $\chi^2=4.1$).
In order to make the fit with the standard interpolation
function compatible with the data of NGC\,7814, the distance to
this galaxy should be increased by a factor $\sim 1.5$.
This is not compatible with the new error determination that allows, at 
$3 \sigma$, an increase of at most $15\%$ \citep{Radburn-Smith+11}.
To improve significatly the fit of NGC\,891 rotation curve the distance should
be decreased by a factor $>2$, incompatible with the new determinations.

\section{Conclusions}

The two galaxies NGC\,891 and NGC\,7814 are representative of two extreme morphologies, with the disk dominating in NGC\,891 and the bulge almost entirely dominating in NGC\,7814.
We have derived new rotation curves for both.
Contrary to previous reports \citep{vdkruit82,vdkruit83}, the shapes of these curves are found to be significantly different. 
They  indicate that in NGC\,7814 the mass is more concentrated to the centre as compared to NGC\,891. 
It resembles, therefore, the distribution of the luminosity, which is more centrally concentrated in NGC\,7814 (bulge) than in NGC\,891 (disk).

A decomposition in bulge, disk and halo shows that in NGC\,891 the disk is the major mass component. 
The bulge contributes about one fourth of the total dynamical mass.
In NGC\,7814 the bulge dominates almost entirely the total luminosity
(90 percent of total).
In the distribution of mass it is undoubtedly the dominant component
in the inner parts. 
In the outer parts the dark matter halo takes over. 
The disk, unless it has an unrealistically high $\ml$ ratio, seems to
be a minor component. 

Standard MOND fits do not work perfectly for both galaxies. 
NGC\,7814 is well fitted by the ``simple'' interpolation function 
\citep{famaey05}.

We conclude that in both galaxies, in their bright optical parts, the
distribution of mass seems to follow closely the distribution of light. 
This implies that either the baryons dominate or the dark matter is
closely coupled to the luminous component.
It would be interesting to repeat the same study on galaxies with 
extreme morphologies such as NGC\,7814 and NGC\,891 but
seen at lower inclination angles.

\begin{acknowledgements}

We thank T.S.\ van Albada, P.C.\ van der Kruit, T.\ Oosterloo, R.\
Peletier, R.\ Sanders, M.\ Verheijen, and E.\ Xilouris 
for helpful comments and stimulating discussions.
We thank C.\ Peng for advice about the fitting of edge-on disks with GALFIT
and D.\ Radburn-Smith for providing the new distances of NGC\,891 and NGC\,7814.
FF is supported by the PRIN-MIUR 2008SPTACC.
PK is supported by the Alexander von Humboldt Foundation.
The Westerbork Synthesis Radio Telescope is operated By ASTRON (Netherlands Institute for Radio Astronomy) with support from the Netherlands Foundation for Scientific Research (NWO).
This work is based in part on observations made with the Spitzer Space Telescope, which is operated by the Jet Propulsion Laboratory, California Institute of Technology under a contract with NASA.
We are grateful to David W. Hogg, Michael R. Blanton, and the Sloan Digital Sky Survey Collaboration for the gri mosaics of NGC\,7814 and to C. Marmo (TERAPIX) for the WIRCam/CFHT multi-color image of NGC\,891.
TERAPIX is funded by the French national research agency (CNRS/INSU), the Programme National de Cosmologie (PNC), the Service d'Astrophysique of the Commissariat l'Energie Atomique (CEA/SAp), the Institut d'Astrophysique de Paris (IAP), and the European FP5 RTD contracts "Astrowise" and "AVO" (Astrophysical Virtual Observatory).
\end{acknowledgements}

\end{document}